\documentclass[hidelinks]{article}
\usepackage{arxiv}
\usepackage{listings}
\usepackage{algorithm}
\usepackage{graphicx}
\usepackage{tikz}[]
\usepackage{tcolorbox}
\usetikzlibrary{positioning, arrows.meta}
\usepackage{enumitem}
\usepackage{multirow}
\usepackage{listings}
\usepackage{makecell}
\usepackage{booktabs}
\usepackage{caption}
\usepackage{subcaption}
\usepackage{algpseudocode}
\usepackage{color}
\usepackage{enumerate}
\usepackage{tabularx}
 \usepackage{underscore}
\definecolor{light-gray}{gray}{0.95}
\usepackage{url}
\usepackage{rotating}
\usepackage{multirow}
\usepackage{amsmath}
\usepackage{tikz}
\usepackage{pgfplots}

\definecolor{codebg}{HTML}{F8F9FA}      
\definecolor{codeframe}{HTML}{D0D0D0}   
\definecolor{codekw}{HTML}{005CC5}      
\definecolor{codecomment}{HTML}{6A9955} 
\definecolor{codestring}{HTML}{DD1144}     
\definecolor{codenumber}{HTML}{A0A1A7}  

\lstdefinestyle{SQLStyle}{
    language=SQL,
    backgroundcolor=\color{codebg},
    basicstyle=\ttfamily\small,
    keywordstyle=\color{codekw}\bfseries,
    commentstyle=\color{codecomment}\itshape,
    stringstyle=\color{codestring},
    numbers=left,
    numberstyle=\tiny\color{codenumber},
    numbersep=10pt,
    frame=single,
    rulecolor=\color{codeframe},
    tabsize=4,
    showstringspaces=false,
    breaklines=true,
    breakatwhitespace=true,
    captionpos=b,
    aboveskip=1em,
    belowskip=1em,
    keepspaces=true,
    morekeywords={BIGINT,BOOLEAN,CHAR,DATE,DECIMAL,DOUBLE,FLOAT,INT,INTEGER,NUMERIC,REAL,SMALLINT,TIME,TIMESTAMP,VARCHAR,TEXT,SERIAL,USER_ATTRIBUTE,RESOURCE_ATTRIBUTE,ENV_ATTRIBUTE,ABAC_RULE,OF,FOR,TO,Allow,Deny,department, designation, type, sensitivity},
}

\lstdefinestyle{CStyle}{
    language=C,
    backgroundcolor=\color{codebg},
    basicstyle=\ttfamily\small,
    keywordstyle=\color{codekw}\bfseries,
    commentstyle=\color{codecomment}\itshape,
    stringstyle=\color{codestring},
    numbers=left,
    numberstyle=\tiny\color{codenumber},
    numbersep=10pt,
    frame=single,
    rulecolor=\color{codeframe},
    tabsize=4,
    showstringspaces=false,
    breaklines=true,
    breakatwhitespace=true,
    captionpos=b,
    aboveskip=1em,
    belowskip=1em,
    keepspaces=true,
    morekeywords={Oid,AclMode,Relation,SysScanDesc,HeapTuple,TupleDesc,ScanKeyData,bool,Datum,text,RTEPermissionInfo,time_t,int16,Form_pg_abac_env_workday}
}
\lstdefinelanguage{Yacc}{
    morekeywords={CreateAbacRuleStmt},
    sensitive=true,
    morecomment=[l]{//},
    morecomment=[s]{/*}{*/},
    morestring=[b]",
}

\lstdefinestyle{YaccStyle}{
    language=Yacc,
    backgroundcolor=\color{codebg},
    basicstyle=\ttfamily\small,
    keywordstyle=\color{codekw}\bfseries,
    commentstyle=\color{codecomment}\itshape,
    stringstyle=\color{codestring},
    numbers=left,
    numberstyle=\tiny\color{codenumber},
    numbersep=10pt,
    frame=single,
    rulecolor=\color{codeframe},
    tabsize=4,
    showstringspaces=false,
    breaklines=true,
    breakatwhitespace=true,
    captionpos=b,
    aboveskip=1em,
    belowskip=1em,
    keepspaces=true,
    alsoletter={\%,\$},
}
\lstdefinestyle{formatStyle}{
    backgroundcolor=\color{codebg},
    basicstyle=\ttfamily\small,
    keywordstyle=\color{codekw}\bfseries,
    commentstyle=\color{codecomment}\itshape,
    stringstyle=\color{codestring},
    numbers=none,
    numbersep=10pt,
    frame=single,
    rulecolor=\color{codeframe},
    tabsize=4,
    showstringspaces=false,
    breaklines=true,
    breakatwhitespace=true,
    captionpos=b,
    aboveskip=1em,
    belowskip=1em,
    keepspaces=true,
}

\AtBeginDocument{
  \providecommand\BibTeX{{%
    \normalfont B\kern-0.5em{\scshape i\kern-0.25em b}\kern-0.8em\TeX}}}

\begin{document}

\title{Generation of Human Comprehensible Access Control Policies from Audit Logs}

\author{%
  Gautam~Kumar\\
  Indian Institute of Technology Kharagpur, India\\
  \texttt{gautam1124bihar@kgpian.iitkgp.ac.in}   \And
  Ravi~Sundaram \\
  Northeastern University, Boston, USA\\
  \texttt{r.sundaram@northeastern.edu} 
   \And
  Shamik~Sural \\
  Indian Institute of Technology Kharagpur, India\\
  \texttt{shamik@cse.iitkgp.ac.in} 
}


\date{}
\renewcommand{\headeright}{}
\renewcommand{\undertitle}{}
\maketitle




\begin{abstract}
Over the years, access control systems have become increasingly more complex, often causing a disconnect between what is envisaged by the stakeholders in decision-making positions and the actual permissions granted as evidenced from access logs. For instance, Attribute-based Access Control (ABAC), which is a flexible yet complex model typically configured by system security officers, can be made understandable to others only when presented at a high level in natural language. Although several algorithms have been proposed in the literature for automatic extraction of ABAC rules from access logs, there is no attempt yet to bridge the semantic gap between the machine-enforceable formal logic and human-centric policy intent. Our work addresses this problem by developing a framework that generates human understandable natural language access control policies from logs. We investigate to what extent the power of Large Language Models (LLMs) can be harnessed to achieve both accuracy and scalability in the process. Named LANTERN (\underline{L}LM-based \underline{A}BAC \underline{N}atural \underline{T}ranslation and \underline{E}xplanation for \underline{R}ule \underline{N}avigation), we have instantiated the framework as a publicly accessible web based application for reproducibility of our results. 

\end{abstract}

\keywords{Attribute-based Access Control (ABAC), Rule Mining, Policy Summarization, Large Language Model (LLM)}


\section{Introduction}
\label{sec:intro}
Security governance and strategic risk management in any organization depend on a proper understanding of access control policies, while system logs provide a ground truth of permissions actually granted to users over various resources \cite{pernuldbseclogs}\cite{accesslogssacmat2025}. However, a common misconception in policy management is that a system's access control logic is always derived from a centrally managed, formal document. In reality, many organizations operate on legacy systems (such as DAC or RBAC) where permissions have evolved organically and semi-autonomously over years. This leads to policy drift, where the permissions actually utilized by users are recorded as the ground truth in system logs which diverge significantly from the high-level security objectives imagined by stakeholders. Consequently, the original policy documentation, is often obsolete or incomplete.  Several algorithms have been developed for extracting policies in a target access control model like Attribute-based Access Control (ABAC) \cite{hu2014nist} from logs - potentially generated by a more traditional access control system like Discretionary Access Control (DAC) or Role-based Access Control (RBAC) \cite{stolleraccesslogmining}\cite{jamesaccesslogmining}\cite{ransam}.

However, in practice, such a policy mining process faces a number of significant barriers. The first among these is the technical overhead of implementing policy mining algorithms, which are complex engines that require custom scripting. A security analyst must be aware of parsing logs and attribute file formats, load the data, and orchestrate the entire mining process. This phase is time consuming and also requires significant programming expertise. In addition, the steps have to be repeated periodically as new log data becomes available. If a new and better algorithm comes up from the research community, it would again require effort as well as time for coding and implementation by the organization. The second deterrent is the opacity of the results, since the output of a policy miner is a set of formal syntactic rules designed for automated enforcement by a machine. These are often complex and not easily understandable by stakeholders responsible for policy governance, such as Managers, Department Chairs, and Deans, who may not be aware of the latest technological developments in access control. This wide knowledge gap between the output of policy mining and human comprehension of the same makes it difficult to validate the mined policies against the stated security objectives of the organization.

To address the above challenges, in this paper we introduce a framework named LANTERN
(\underline{L}LM-based \underline{A}BAC \underline{N}atural \underline{T}ranslation and \underline{E}xplanation for \underline{R}ule \underline{N}avigation). It generates a natural language abstraction of ABAC rules derived from access logs and prior knowledge of attribute values of users, resources, and environment - the three entity types used in ABAC. LANTERN harnesses the reasoning capabilities of Large Language Models (LLMs) to synthesize semantically faithful yet human-readable policy statements. The framework integrates symbolic policy inferencing with LLM-driven linguistic transformation to ensure accuracy, interpretability, and scalability. 
Note that the goal of LANTERN differs from methods that generate Machine-Enforceable Security Policies (MESPs) from Natural Language Access Control Policies (NLACPs) as proposed in \cite{manar2}\cite{indrakshisacmat23}\cite{miansacmat2025}\cite{sonune2025lmntoolgeneratingmachine}. 
Results of an extensive set of experiments demonstrate that LANTERN can achieve high policy fidelity while significantly improving interpretability.
and alignment with stakeholder intent. 
We share the link to a free web-based version of LANTERN for reproducibility of our results. Further, the entire source code has being made publicly available. 
The key contributions of this paper are summarized below.

\begin{itemize}
\item Human-Readable Policy Synthesis: We introduce the first framework that generates natural-language access control policies from access logs and entity attribute values, making mined policies interpretable to decision-makers.
\item Hybrid Architecture: LANTERN combines symbolic rule extraction with LLM-based linguistic transformation, ensuring fidelity to the underlying access semantics while maintaining grammatical and conceptual clarity.
\item Prompt-Engineered Accuracy and Scalability: We design a structured prompting and chunking strategy that enables LANTERN to handle variably-sized logs and attribute datasets while preserving inference accuracy.
\item Extensive Empirical Evaluation: Using several synthetic data-sets, we demonstrate the effectiveness of LANTERN in achieving its intended goal.
\item Open Deployment: We release a web-based version 
of LANTERN for enabling interacting with the tool.
\end{itemize}


The rest of the paper is organized as follows. We introduce some of the foundational concepts in Section \ref{sec:prelims}. The design considerations as well as implementation details are presented in Section \ref{sec:lanterndetails}, followed by experimental results in Section \ref{sec:evaluation}. Related work is reviewed in Section \ref{sec:related} and finally we draw conclusions, identify shortcomings, and provide future directions in Section \ref{sec:concl}.

\section{Preliminaries}
\label{sec:prelims}
ABAC is a dynamic model of access control that has received significant attention in information system security. It is built on the notion of attributes which are characteristics of entities. There are primarily three types of entities, namely, subjects or users, resources or objects, and environmental conditions \cite{hu2014nist}. 
Each of these entities is characterized by a set of attributes, i.e., user attributes, object attributes and environmental attributes. The activities that a particular user can perform on a particular object in a specific environmental condition are called operations. Every user, resource, and environment entity is assigned a value for each of the corresponding attributes. 
Another, and possibly the most important, component of ABAC is a Policy, which is a set of authorization rules that govern who can access what object under what environmental conditions. 

For enabling ABAC in any organization, the above-mentioned components need to be appropriately defined or derived from existing data sources - a step commonly known as Policy Mining. 
The necessity of policy mining arises primarily during system migration or security audits. When an organization transitions from a legacy model to ABAC, they often lack a formal attribute-based rules. In such cases, the system logs serve as the only reliable repository of existing access patterns. The policy mining problem thus takes as input a set of users along with their attribute values (e.g., department, designation, role), a set of resources and their attribute values (e.g., type, sensitivity, owner), a set of environmental conditions and their attribute values (e.g., time of day, request originating subnet), along with a collection of prior access decisions—typically recorded as logs containing tuples of the form $\langle$user, resource, operation, timestamp$\rangle$. The desired output is a set of ABAC authorization rules as mentioned above.

The challenge in ABAC policy mining lies in finding a minimal set of concise rules that grant the observed permissions, while avoiding excessive over-assignments or under-assignments. A well-mined policy should be both precise (covering all of the logged access patterns but no additional ones) and simple (containing the smallest possible set of rules, each with low structural complexity). ABAC policy mining, which is known to be an NP-Complete problem, has been an active research area for over a decade. One of the early attempts is due to Xu and Stoller~\cite{tdscstoller}, who proposed an algorithm for mining policies from logs, establishing key concepts like seed-based generalization and the Weighted Structural Complexity (WSC) metric to measure policy quality. Since then, policy mining work has been extended in various directions: mining policies that include both permit and deny rules~\cite{iyer2018mining}, developing algorithms for handling incomplete or sparse logs~\cite{abu2020mining}, applying machine learning techniques like clustering~\cite{shang2024abac} and evolutionary algorithms~\cite{medvet2015evolutionary}, and formulating mining as a functional dependency discovery problem~\cite{talukdar2017efficient}. There are also several recent approaches that consider practical system level constraints in ABAC policy mining \cite{ransam}\cite{gunjantoit}\cite{contemporaneous}.

\begin{figure*}[t]
  \centering
  \label{fig:two_phase_process}
  \includegraphics[width=\textwidth]{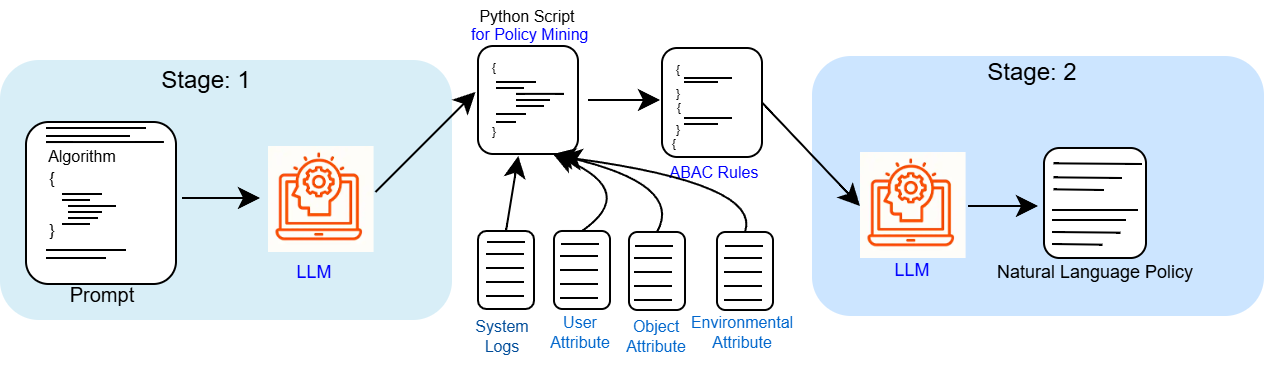}
  \caption{Two-stage, LLM-driven architecture of LANTERN: Stage 1 (Code Generation) and Stage 2 (Policy Summarization)}
  \label{fig:architecture}
\end{figure*}
\section{LANTERN Design and Implementation}
\label{sec:lanterndetails}

In this section, we present the detailed design of LANTERN.
The main features of implementation are also highlighted. LANTERN design and implementation attempts to address the following four research questions.\\
  \textbf{ RQ1:} Can LANTERN produce working integration scripts with the generated code successfully handling diverse data formats?\\
  \textbf{RQ2:} Does the mining code generated by the LLM component of LANTERN successfully discover policies from both native ABAC logs and legacy DAC logs?\\
  \textbf{RQ3:} Can LLM-based policy summarization transform technical rules into comprehensible business documentation?\\
  \textbf{RQ4:} What is a minimalist deign of LANTERN that can achieve all the objectives outlined in Section \ref{sec:intro}?

Through our experiments detailed in Section \ref{sec:evaluation} we will address the above questions.

\subsection{Design Considerations for LANTERN}
\label{subsec:lanterndesign}
LANTERN is an LLM-powered framework that transforms the traditionally expert-dependent process of ABAC policy mining \cite{stolleraccesslogmining} into an accessible, conversational workflow. It operates through a two-stage pipeline as shown in Figure~\ref{fig:architecture} designed to address the dual barriers of implementation overhead and result incomprehensibility. Crucially, this pipeline treats system logs as a behavioral footprint of the organization. Because logs record actual user interactions which includes manual changes or old settings that were never officially documented, they represent the \textit{de facto} policy currently in use. In the first stage of LANTERN, an LLM generates deployment ready code for mining ABAC rules from system logs. The code is executed locally to generate the rules, which typically consist of complex attribute expressions, constraints, and set notations. In the second stage, the rules which are mostly incomprehensible to non-experts, are converted into executive-level natural language summaries tailored to different stakeholder groups. This process turns messy behavioral data into a readable policy summary that managers can understand, verify and correct, to match their intended security goals. This end-to-end pipeline represents a paradigm shift in policy mining, replacing a workflow that previously required programming expertise, ABAC knowledge and months of effort, with a streamlined process achievable through simple data uploads and natural language interactions. 
The main components of LANTERN are described in the following sub-sections.

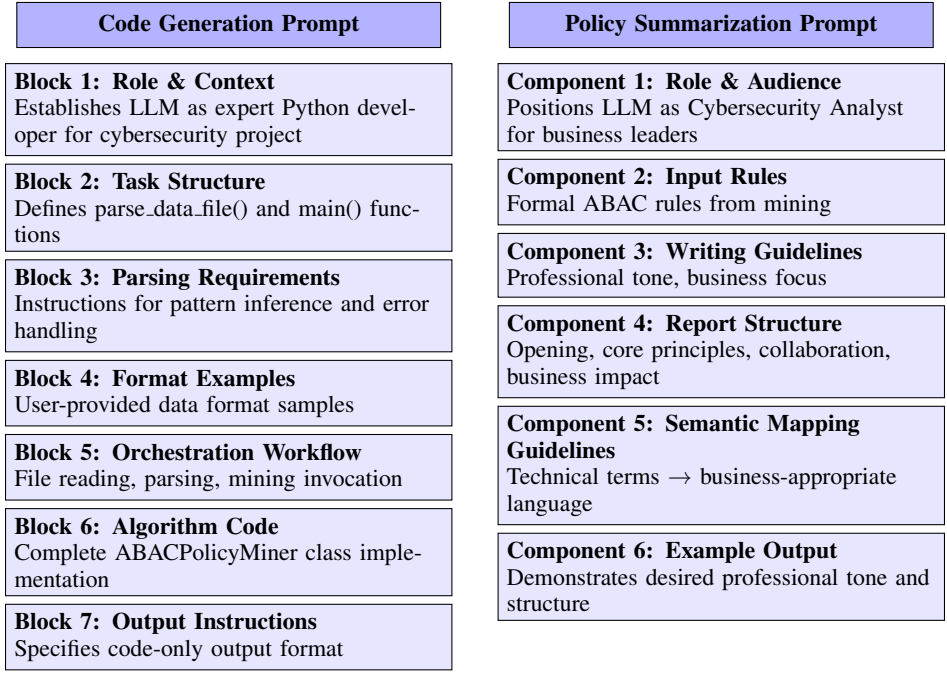
\begin{figure*}[h]
\centering
\begin{tikzpicture}[
    node distance=0.3cm,
    block/.style={rectangle, draw, fill=blue!10, text width=5.7cm, align=left, minimum height=0.6cm, font=\small},
    title/.style={rectangle, draw, fill=blue!30, text width=5.4cm, align=center, minimum height=0.6cm, font=\small\bfseries}
]
\node[title] (cg_title) {Code Generation Prompt};
\node[block, below=0.2cm of cg_title] (cg1) {\textbf{Block 1: Role \& Context}\\
Establishes LLM as expert Python developer for cybersecurity project};
\node[block, below=0.1cm of cg1] (cg2) {\textbf{Block 2: Task Structure}\\
Defines parse\_data\_file() and main() functions};
\node[block, below=0.1cm of cg2] (cg3) {\textbf{Block 3: Parsing Requirements}\\
Instructions for pattern inference and error handling};
\node[block, below=0.1cm of cg3] (cg4) {\textbf{Block 4: Format Examples}\\
User-provided data format samples};
\node[block, below=0.1cm of cg4] (cg5) {\textbf{Block 5: Orchestration Workflow}\\
File reading, parsing, mining invocation};
\node[block, below=0.1cm of cg5] (cg6) {\textbf{Block 6: Algorithm Code}\\
Complete ABACPolicyMiner class implementation};
\node[block, below=0.1cm of cg6] (cg7) {\textbf{Block 7: Output Instructions}\\
Specifies code-only output format};

\node[title, right=0.9cm of cg_title] (ps_title) {Policy Summarization Prompt};
\node[block, below=0.2cm of ps_title] (ps1) {\textbf{Component 1: Role \& Audience}\\
Positions LLM as Cybersecurity Analyst for business leaders};
\node[block, below=0.1cm of ps1] (ps2) {\textbf{Component 2: Input Rules}\\
Formal ABAC rules from mining};
\node[block, below=0.1cm of ps2] (ps3) {\textbf{Component 3: Writing Guidelines}\\
Professional tone, business focus};
\node[block, below=0.1cm of ps3] (ps4) {\textbf{Component 4: Report Structure}\\
Opening, core principles, collaboration, business impact};
\node[block, below=0.1cm of ps4] (ps5) {\textbf{Component 5: Semantic Mapping Guidelines}\\
Technical terms → business-appropriate language};
\node[block, below=0.1cm of ps5] (ps6) {\textbf{Component 6: Example Output}\\
Demonstrates desired professional tone and structure};



\end{tikzpicture}
\caption{Structured prompt architecture for LANTERN: Code generation (left) and Policy Summarization (right)}
\label{fig:prompt-structure}
\end{figure*}


\textbf{Code Generation using LLM:} This stage, as shown on the left side of Figure \ref{fig:prompt-structure}, addresses the implementation barrier through automated code generation with custom data parser. The design is format-agnostic, meaning it can handle diverse and inconsistent data representations without requiring users to provide formal schema definitions or structural annotations. The user simply uploads a representative snapshot of their data (the \textit{structure}) for users, resources, and logs. LANTERN then acts as an intelligent translator. It studies these snapshots to identify how the information is organized (e.g., recognizing column names, brackets, or commas). It generates a custom Python parser specifically for those files. This flexibility is a major advantage over traditional tools, for example, if a log file contains a mix of different styles, like \texttt{<alice department:Finance designation:Manager>} and \texttt{alice,Finance,Manager} for users. By recognizing these structural patterns from raw snapshots, LANTERN automatically 
generates the required Python parser, effectively removing the technical bottleneck of manual script development.

The design of LANTERN deliberately constrains the task of the LLM to code synthesis rather than algorithmic invention, ensuring that the generated code is deterministic and testable while avoiding hallucination risks in logical reasoning.
The main design idea is to treat format specification as a demonstration task instead of a description task. Rather than asking users to explain format rules with technical terms, the system lets them provide simple examples. Such type of design leverages the pattern recognition strengths of the LLM from training on millions of parsing code examples. This format-agnostic approach ensures that LANTERN works well for access records with different conventions as the system adapts to formats rather than requiring format standardization.


\textbf{LLM-Based Policy Translation:} This second stage addresses comprehension barriers through contextual summarization as shown in the right side of Figure \ref{fig:prompt-structure}. The design positions the LLM as a multi-stakeholder translator that transforms individual formal rules into distinct explanations optimized for different audiences: concise executive summaries for approval workflows, detailed compliance documentation for auditors, and actionable guidance for policy administrators. The architectural design provides a rich context alongside each formal rule, including the organization's attribute schema, rule coverage statistics, and optional policy guidelines. 
The design specifies multi-level explanation structures containing one-sentence summaries, detailed business explanations, concrete affected-user examples, security assessments, and compliance notes, thus ensuring single-pass generation serves diverse stakeholder needs simultaneously. The critical design choice is structured prompting that explicitly prevents ABAC jargon while requiring business-focused explanations.

\subsection{Implementation Details}
\label{subsec:LANTERN_impl}
We have instantiated the LANTERN framework as described in the previous sub-section as a modular Flask based web application\footnote{\url{https://lanterntoolkit.pythonanywhere.com/}}. 
The complete source code is also available on our GitHub link\footnote{\url{https://github.com/LANTERN-toolkit/LANTERN}}.
Detailed considerations for implementing LANTERN are described below. 

\textbf{Prompt Engineering for Code Generation:} The code generation workflow begins when a user submits format examples through the web interface of LANTERN. The implementation constructs a prompt organized into logically structured blocks (Blocks 1-7 on the left side of Figure \ref{fig:prompt-structure}), each serving a distinct purpose in guiding the LLM through the generation task. The first block establishes role and context, 
which activates relevant domain knowledge and coding conventions, improving output quality through appropriate priming. The block explicitly identifies the core system component — the $ABACPolicyMiner$ class that the generated code must integrate with, establishing the technical context for the entire task. The second block tells the LLM to create two separate functions: one called $parse\_data\_file$ that reads and interprets the data files, and another called $main$ that runs the whole process from start to finish. Breaking the task into two pieces helps the LLM write cleaner, more organized code instead of one monolithic program.

The third block gives instructions for the parsing function. Instead of telling the LLM exactly how to parse the data, it says, ``Look at the examples and figure out the pattern yourself''. This lets the LLM use its pattern recognition abilities, quite the same way a human programmer would study unfamiliar data to understand its structure. The block also tells LLM to handle problems gracefully: skip lines that start with \# (comments), ignore blank lines, and print warning messages for data that does not match the expected pattern instead of crashing.

The fourth block inserts the actual format examples that users provide. These examples show the LLM real patterns to learn from, like how fields are separated, what wrapper characters are used, and how attributes are structured. The fifth block describes what the main function should do: read the three data files (and handle missing file errors), call the parsing function to convert raw text into structured data, create an instance of the mining class, run the mining process, and display the results in a readable format.

The algorithm block is a critical component, where the prompt includes the complete source code of both the $ABACRule$ and $ABACPolicyMiner$ classes spanning several hundred lines of Python code. Providing the full mining algorithm implementation is essential for three reasons. First, it gives the LLM precise knowledge of the exact data structures expected by the mining engine, ensuring that the generated parsing code produces compatible output formats. Second, by showing the complete algorithm, we constrain the LLM task to generating integration code rather than inventing mining logic, thereby reducing hallucination risks since the LLM never attempts to create policy rules or mining heuristics itself. Third, it enables the LLM to verify compatibility by understanding how its parsing code output flows into the input interface of the algorithm, catching structural mismatches during generation rather than at runtime.

The final block contains explicit output instructions: ``Provide ONLY the complete, self-contained Python code for the $parse\_data\_file$ function, including necessary imports. Do not include any example usage or explanations''. This directive prevents the LLM from adding explanatory prose that would require post-processing.

\textbf{Mining Algorithm Implementation:} The LANTERN architecture provides flexibility in mining algorithm selection. Organizations can employ any ABAC mining approach while retaining the LLM-powered accessibility benefits. Our current implementation uses an improved version of the seed-based algorithm proposed in ~\cite{stolleraccesslogmining}, chosen for its proven effectiveness and extensive validation in prior research. However, this choice is only incidental to the value proposition of LANTERN. Organizations can substitute new and improved machine learning approaches, clustering techniques, or custom algorithms by replacing the $ABACPolicyMiner$ class, while preserving all LLM-generated components.

The comprehensive improvements done by us over~\cite{stolleraccesslogmining} address schema independence to ensure consistent performance across diverse attribute configurations. Traditional policy mining exhibits bias when users and resources have different attribute counts, as quality metrics implicitly assume balanced schemas. Our key enhancements include effective pattern tracking that captures not only user-resource correlations but also patterns within user attributes and resource attributes, enabling rule discovery even when one entity type has minimal attributes. We employ side-normalized quality metrics that average out user and resource contributions rather than summing them, thus preventing attribute count differences from distorting rule scores. An entropy-based bucketing is also implemented that selects discriminative attributes based on information gain rather than position, ensuring semantically meaningful attributes drive rule generation regardless of schema structure. These modifications maintain the core seed-generalization-refinement approach, while eliminating schema biases.


\textbf{Policy Summarization Implementation:} The summarization workflow shown on the right side of Figure \ref{fig:prompt-structure} is comprised of five structured components that guide the LLM through professional policy documentation generation.
We employ Gemini 2.5 Pro\footnote{\url{https://aistudio.google.com/}} as the LLM for this translation task due to its superior performance on nuanced natural language generation and audience adaptation capabilities. 

The first component establishes role and audience context, positioning the LLM as a \textit{Cybersecurity Analyst tasked with communicating complex security policies to business leaders}. Such a framing activates domain-appropriate knowledge about professional communication styles, risk management terminology, and executive stakeholder expectations. The second component takes the ABAC rules as input which was obtained as output of Stage 1. The third component provides writing guidelines that specify tone (professional and authoritative), content focus (explaining business logic and security principles rather than technical specifications), and clarity requirements (connecting access controls to job functions and responsibilities).

The fourth component mandates a structured report format comprising: (i) an opening statement summarizing policy purpose and principles, (ii) core access principles organized under clear headings such as Role-based Access, Data Sensitivity, and Principle of Least Privilege, (iii) cross-functional collaboration explanations showing secure data sharing mechanisms, and (iv) a business impact conclusion emphasizing asset protection and operational support. This structure ensures consistency across generated summaries while maintaining executive readability through scannable organization.

The fifth component provides specific linguistic transformation instructions to ensure the final output is easily understandable to non-technical stakeholders. We explicitly instructed the LLM to simplify and map technical system jargon to high-level business descriptors. To guide the model in this process, we provided a set of few-shot examples for instance, \textit{technical rules} are transformed into \textit{organizational policies}, \textit{safe systems} are described as \textit{protected assests}. The prompt includes example outputs demonstrating the desired professional tone, showing how formal rule tuples should transform into cohesive governance narratives.

\begin{figure*}[t]
    \centering
    
    \begin{subfigure}{0.32\textwidth}
        \includegraphics[width=\linewidth]{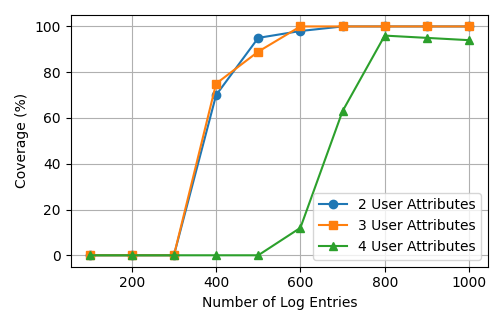}
        \caption{Coverage vs. User Attributes.}
        \label{fig:coverage_user_attr}
    \end{subfigure}
    \hfill 
    \begin{subfigure}{0.32\textwidth}
        \includegraphics[width=\linewidth]{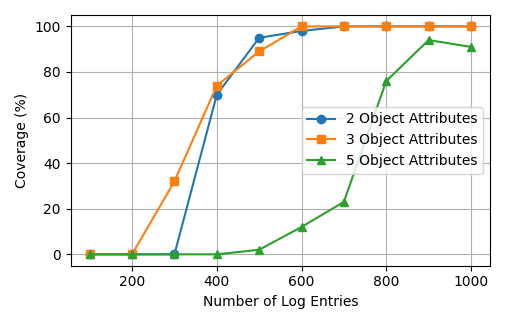}
        \caption{Coverage vs. Object Attributes.}
        \label{fig:coverage_object_attr}
    \end{subfigure}
    \hfill 
    \begin{subfigure}{0.32\textwidth}
        \includegraphics[width=\linewidth]{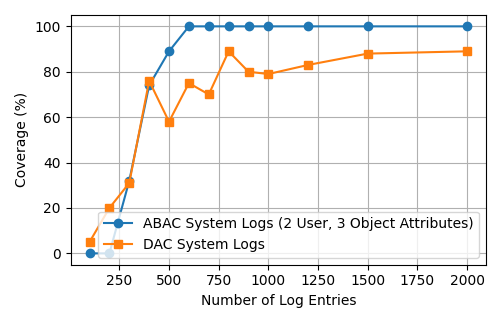}
        \caption{ABAC vs. DAC Coverage.}
        \label{fig:coverage_abac_dac}
    \end{subfigure}
    
    \caption{Evaluation of policy mining coverage. (a) Impact of increasing user attributes for ABAC logs (b) Impact of increasing object attributes  for ABAC logs and (c) Effectiveness on logs from native ABAC and legacy DAC systems.}
    \label{fig:coverage_analysis_row}
\end{figure*}

\begin{figure*}[t]
    \centering
    
    \begin{subfigure}{0.32\textwidth}
        \includegraphics[width=\linewidth]{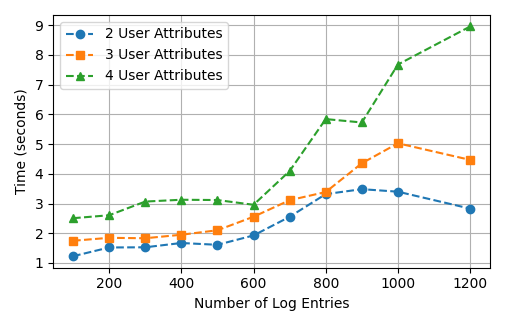}
        \caption{Time vs. User Attributes.}
        \label{fig:time_user_attr}
    \end{subfigure}
    \hfill 
    \begin{subfigure}{0.32\textwidth}
        \includegraphics[width=\linewidth]{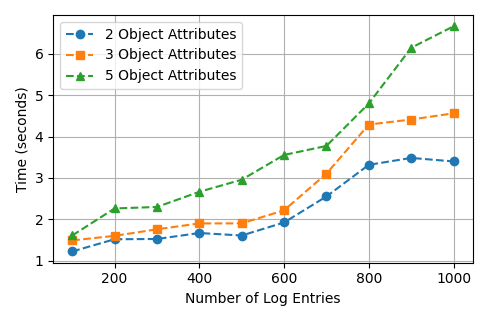}
        \caption{Time vs. Object Attributes.}
        \label{fig:time_object_attr}
    \end{subfigure}
    \hfill 
    \begin{subfigure}{0.32\textwidth}
        \includegraphics[width=\linewidth]{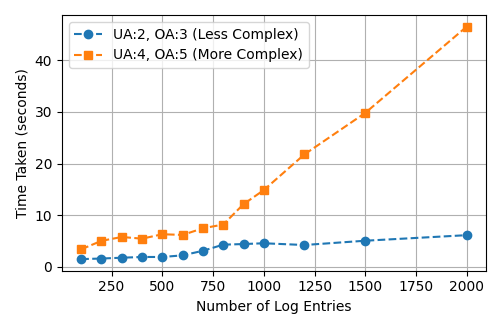}
        \caption{Time: Low vs. High Complexity.}
        \label{fig:time_complexity_comp}
    \end{subfigure}
    
    \caption{Evaluation of policy mining performance in terms of execution time in seconds for ABAC logs}
    \label{fig:performance_analysis_row}
\end{figure*}

\section{Experimental Results}
\label{sec:evaluation}

In the following sub-sections, we first describe our data generation step, followed by evaluation of Stages 1 and 2 of LANTERN, and finally carry out ablation studies.

\begin{table}[t]
  \caption{Summary of synthetic dataset parameters}
  \label{tab:dataset_params_accurate}
  \centering
  \small
  \begin{tabular}{p{0.18\textwidth} p{0.41\textwidth}
  p{0.41\textwidth}}
    \toprule
    \textbf{Dataset} & \textbf{User and Object Attributes} & \\
    \midrule

    \textbf{ABAC} &
    \textbf{User Attributes:}
    \begin{itemize}[leftmargin=*, noitemsep, topsep=0pt]
        \item \texttt{department} (10 values)
        \item \texttt{designation} (8 values)
        \item \texttt{clearance} (3 values)
        \item \texttt{region} (3 values)
    \end{itemize} &
    \textbf{Object Attributes:}
    \begin{itemize}[leftmargin=*, noitemsep, topsep=0pt]
        \item \texttt{type} (4 values)
        \item \texttt{sensitivity} (3 values)
        \item \texttt{region} (3 values)
        \item \texttt{project} (3 values)
        \item \texttt{owner\_dept} (4 values)
    \end{itemize}
    \\

    \midrule

    \textbf{DAC} &
    \textbf{User Attributes:}
    \begin{itemize}[leftmargin=*, noitemsep, topsep=0pt]
        \item \texttt{department} (10 values)
        \item \texttt{designation} (31 values)
    \end{itemize} &
    \textbf{Object Attributes:}
    \begin{itemize}[leftmargin=*, noitemsep, topsep=0pt]
        \item \texttt{type} (9 values)
        \item \texttt{sensitivity} (4 values)
    \end{itemize}
    \\
    \bottomrule
  \end{tabular}
\end{table}

\subsection{Dataset Generation}
\label{subsec:dataset}

To carry out controlled evaluation, we generated two distinct synthetic datasets as described below. While both datasets simulate an organization of the same scale, they are governed by different access control models. The key parameters defining the scale of our synthetic organization are summarized in Table~\ref{tab:dataset_params_accurate}. 

\textbf{ABAC Dataset:} Our primary dataset simulates an organization governed by a complex, attribute-based policy containing 12 distinct multi-attribute rules in its full configuration (4 user attributes, 5 object attributes). To evaluate scalability and complexity handling, we created multiple schema variants (U2/O2, U3/O2, U4/O5), where simpler configurations use proper subsets of the master policy. For each configuration, we generated log files of varying sizes (100 to 2000 entries) by creating random user-resource-action tuples and querying the ground truth policy. To ensure dense positive examples, we recorded all \textit{Allow} decisions while sampling \textit{Deny} decisions with 1\% probability.

\textbf{DAC Dataset:} To test LANTERN on logs from non-ABAC systems, we generated a second dataset simulating a larger organization (74 users, 60 objects) governed by DAC through per-object Access Control Lists (ACLs). Critically, these ACLs were not randomly generated but programmatically constructed to embed realistic attribute-based patterns: departmental ownership (Finance objects owned by Finance users), role-based permissions (senior roles receiving broader access), cross-departmental workflows (Sales accessing Marketing objects for collaboration), and sensitivity constraints (Confidential objects with minimal ACLs). This creates an identity-driven dataset with discoverable attribute correlations simulating the real-world challenge of migrating legacy DAC systems to ABAC. Log generation targeted a precise 90\% \textit{Allow} and 10\% \textit{Deny} ratio for consistent evaluation conditions.

\subsection{Evaluation of LLM-Generated Code}
\label{subsec:evalLLMgencode}
The first set of experiments investigates whether LLM-powered code generation of LANTERN can successfully automate deployment across diverse organizational data formats. Towards this, 
we evaluated the ability of the LLM code generator to produce correct Python integration scripts from format specifications. Our test suite was comprised of three realistic format variations that organizations might encounter. (i) Format 1 - Angle Bracket with Colons: Standard format where lines are wrapped in $\langle$...$\rangle$ with space-delimited `key:value` pairs. (ii) Format 2 - CSV Format: No wrapper characters, comma-separated values. (iii) Format 3 - Pipe-Delimited: Vertical bars separating fields.

For each format, we provided example snippets to the LLM code generator along with the complete mining algorithm implementation (\textit{ABACPolicyMiner} class) in the algorithm block of the prompt. The generator produced complete Python scripts including the \textit{parse\_data\_file}() function tailored to each format and the \textit{main}() orchestration function. Each LLM-generated script was executed on sample data files matching the corresponding format. The validation criteria were:

\begin{itemize}
    \item \textbf{Syntactic Correctness:} Generated code must be valid Python without syntax errors
    \item \textbf{Format Compatibility:} Parsing functions must correctly extract entities and attributes from the specified format
    \item \textbf{Interface Compatibility:} Generated code must produce data structures matching the mining algorithm's expected input (dictionaries for users/resources, list for logs)
    \item \textbf{Execution Completeness:} Scripts must successfully complete the full workflow from file reading through policy generation
\end{itemize}

For the codes generated by LLM, all the above criteria were satisfied. Hence, research question RQ1 is successfully answered. 


Having validated that the LLM generates working integration code, we next evaluate whether the complete pipeline (LLM generated parsing + provided mining algorithm) successfully discovers the syntactic ABAC rules. 
Figures~\ref{fig:coverage_user_attr} and~\ref{fig:coverage_object_attr} demonstrate the effectiveness of the LLM-generated code when integrated with the mining algorithm. For all tested configurations, the system exhibits clear learning curves: permission coverage starts low on sparse logs but rapidly improves with data density, consistently achieving 100\% coverage once approximately 600-800 log entries are available. This confirms that the LLM-generated parsing code correctly extracts and transforms data, enabling the mining algorithm to perfectly reconstruct ground truth policies. We also observed (omitted here due to page limit) that as coverage reaches 100\%, number of rules decreases, demonstrating that the complete pipeline (LLM integration + mining algorithm) produces concise, optimal policies rather than overly-specific rule lists.

Figure~\ref{fig:coverage_abac_dac} shows system performance on the DAC dataset and compares it with the ABAC dataset. The results reveal a phase transition: performance is poor below 400 entries but rapidly improves on denser logs, stabilizing at 80-89\% coverage for logs with 800-2000 entries. This plateau is not a failure but rather validates that the LLM-generated code successfully enables the mining algorithm to identify core attribute-driven business logic (departmental ownership, role hierarchies) embedded in ACLs, while appropriately ignoring the 10-20\% of truly discretionary permissions with no repeating patterns. It demonstrates that LANTERN effectively handles real-world challenges of policy migration from legacy systems.

Figures~\ref{fig:performance_analysis_row}(a)-(c) characterize the computational performance of Stage 1 of LANTERN. Execution time grows slowly with attribute dimensionality, ranging from under 10 seconds for simple configurations (U2/O2) to about 46 seconds for complex scenarios (U4/O5 with 2000 entries). This trend reflects the combinatorial search space expansion in the association rule mining phase of the mining algorithm. While the LLM-generated code introduces minimal overhead (parsing and data transformation complete in milliseconds), these results indicate that the computational complexity of the mining algorithm remains the bottleneck at high attribute counts—a characteristic inherited from the underlying mining approach (Recollect from Section \ref{sec:prelims} that mining of optimal ABAC rules is NP-Complete), not a limitation of the LLM-generated integration layer. The results depicted in Figures \ref{fig:coverage_analysis_row} and  \ref{fig:performance_analysis_row} together answer research question RQ2.

To further validate the scalability and the real-world effectiveness of the pipeline, i.e., to answer research question  RQ2, we also evaluated its performance against five widely-used, large-scale benchmark datasets from the access control literature \cite{tdscstoller}. As summarized in Table \ref{tab:benchmark_results}, the results confirm the scalability of LANTERN pipeline, which maintains excellent performance even when tested against large and complex log files.
\begin{table}[H]
\centering
\caption{LANTERN Performance on Benchmark Datasets.}
\label{tab:benchmark_results}
\begin{tabularx}{\linewidth}{X r r r} 
\toprule
\textbf{Benchmark Dataset} & \textbf{Log Size} & \textbf{Coverage (\%)} & \textbf{Time (s)} \\
\midrule
Healthcare Dataset 0 - lg &  18k & 100 & 355 \\
Healthcare Dataset 1 - lg &  18k & 98 & 218 \\
Healthcare Dataset 2 - sm &   2k  & 100 & 46  \\
University Dataset 0 - lg &  16k & 99 & 431 \\
University Dataset 1 - sm &   4k & 97 &  70 \\
\bottomrule
\end{tabularx}
\end{table}
\subsection{Evaluation of Comprehensibility of Generated Policies}
\label{subsec:evalLLMgensummary}

The second stage of our evaluation assesses whether LLM-powered summarization of LANTERN successfully bridges the comprehension gap between technical policy artifacts and business stakeholders. The summarizer effectiveness depends on the quality of prompt engineering. Our prompt is designed not to just to ask for a summary, but to provide a specific \textit{persona} and a set of instructions that guide the LLM toward producing a professional, clear, and business-focused output. 

To systematically assess the quality and accuracy of generated natural language policies, we developed an automated verification pipeline using Claude 3.5 Sonnet\footnote{\url{https://www.anthropic.com/news/claude-3-5-sonnet}} as an independent evaluator (LANTERN uses Gemini for generating the policies). This LLM-driven verification approach enables a scalable evaluation of policies generated by LLM. The verification process operates as follows: For each policy set, both the formal ABAC rules and the generated natural language policy are provided to Claude. The model is explicitly instructed to function as a verifier, analyzing how accurately and comprehensively the natural language description captures each formal rule. The verifier evaluates each of those rules independently against the generated text across multiple dimensions.
The overall accuracy is computed as the mean of individual rule scores. Such an approach enables consistent, reproducible evaluation across multiple policy sets and provides detailed diagnostic information about specific deficiencies in generated policies.

We evaluated our  approach across five policy sets of varying complexity, ranging from 3 to 29 rules. Table~\ref{tab:nl_evaluation} presents the automated verification results. The results demonstrate strong performance across all policy sets, with average accuracies ranging from 90.0\% to 99.8\%. For moderately complex policies (8-22 rules), we achieved near-perfect accuracy (96-99.8\%), with the majority of rules (75-95\%) receiving perfect 100\% scores from the automated verifier. The 3-rule policy set showed lower accuracy (90\%) due to a rule conflation issue where two overlapping rules were incorrectly merged in the natural language output. The 29-rule policy, despite its complexity, maintained 99\% accuracy with 90\% of rules perfectly captured (100\% score) and an additional 10\% achieving excellent scores (95-99\%). Research question RQ3 is thus addressed.

\begin{table}[t]
\centering
\caption{Natural language policy generation accuracy}
\label{tab:nl_evaluation}
\begin{tabular}{ccccc}
\hline
\textbf{\#Rules} &
\makecell{\textbf{Avg. Accuracy}\\\textbf{(\%)}} &
\makecell{\textbf{Perfect}\\\textbf{(100\%)}} &
\makecell{\textbf{Excellent}\\\textbf{(95--99\%)}} \\
\hline
3  & 90.0  & 2 & 0  \\
8  & 96.0  & 6 & 1 \\
17 & 99.1  & 14 & 3 \\
22 & 99.8  & 21 & 1 \\
29 & 99.0  & 26 & 3 \\
\hline
\end{tabular}
\end{table}

\subsection{Ablation Study}
\label{subsec:ablation}
To study the usefulness of our two-stage pipeline design, we conducted an ablation study for analyzing the performance of alternative, simpler approaches. This study traces our research path and provides a clear justification why a hybrid system combining a formal mining algorithm with a constrained LLM is necessary to solve the problem effectively. We evaluated three distinct methodologies: a pure end-to-end LLM approach, a compression-enhanced LLM approach, and our final two-stage hybrid pipeline.

\textit{Alternative 1: End-to-End LLM Generation.}
We first explore the most straightforward approach, i.e., providing the raw access logs directly to an LLM and prompting it to generate both the formal ABAC rules and a natural language summary. This method revealed a significant scalability issue. As shown in Table~\ref{tab:ablation_comparison}, while it performed reasonably well on very small log files ($\le$400 entries), its performance degraded dramatically as the log size increased. Beyond approximately 500 entries, the model's ability to reason over the growing number of entities and attribute relationships collapsed, leading to a sharp decline in log coverage and highly inconsistent results. This failure is primarily attributable to the fixed context window of LLM and its inherent difficulty in performing the complex, multi-step reasoning required for policy synthesis from raw data.

\textit{Alternative 2: Compression-Enhanced LLM.}
To mitigate the scalability issues, we hypothesized that the input size could be reduced by removing redundant log entries. We developed a log compression technique that consolidates repeated permissions (multiple log entries for the same user type, i.e., having same attribute values, resource type, and operation) into a single unique entry. The compressed log was then provided to the LLM. This approach moderately improved performance, allowing the model to handle logs of up to around 700 entries before performance began to degrade. However, as shown in Table~\ref{tab:ablation_comparison}, this was only an incremental improvement. The number of \textit{unique} permissions in a log file still grows significantly, and this compression step was ultimately insufficient to overcome the core limitation of LLM in synthesizing a globally consistent policy from a large set of discrete facts.

\textit{Alternative 3: LLM + Mining Script} (LANTERN). From the limitations of the pure LLM approaches,
we concluded that while LLMs are excellent at language tasks, they are not a substitute for a deterministic algorithm for the core synthesis problem. In this final approach, we use our formal \textit{ABACPolicyMiner} algorithm to handle the scalable and complex task of rule discovery. The LLM is then employed for two specific, constrained tasks it excels at: generating the environment-specific parsing code and summarizing the final, structured rules.

\begin{table}[t]
  \centering
  \caption{Comparison of methodologies for ABAC rule generation across varying log sizes. Note: * denotes LLM execution timed out and no output generated.} 
  \label{tab:ablation_comparison}
  \begin{tabular}{@{}l|cc|cc|cc@{}}
    \toprule
    \textbf{Log} &
    \multicolumn{2}{c|}{\textbf{Pure LLM}} &
    \multicolumn{2}{c|}{\textbf{Comp. + LLM}} &
    \multicolumn{2}{c}{\textbf{LLM + Script}} \\
    \cmidrule(lr){2-3} \cmidrule(lr){4-5} \cmidrule(lr){6-7}
    & \makecell{\textbf{\# Rules}} & \makecell{\textbf{Time} \textbf{(s)}} 
    & \makecell{\textbf{\# Rules}} & \makecell{\textbf{Time} \textbf{(s)}} 
    & \makecell{\textbf{\# Rules}} & \makecell{\textbf{Time} \textbf{(s)}} \\
    \midrule
    400  & 13 & 195 & 29 & 228 & 5  & 1.7 \\
    500  & 10 & 86  & 16 & 105 & 10 & 1.6 \\
    600  & 12 & 170 & 18 & 190 & 13 & 1.9 \\
    800  & 50 & 440 & 20 & 350 & 11 & 3.3 \\
    1000 & 18 & 243 & 15 & 280 & 9  & 3.4 \\
    2000 & 21 & 421 & 18 & 450 & 10 & 17.9 \\
    5000 & 0* & 631 & 11 & 531 & 3 & 19.7 \\
    10000 & 0* & - & 0* & - & 2 & 11.9 \\
    \bottomrule
  \end{tabular}
\end{table}

\begin{figure}[t]
  \centering
  \includegraphics[width=0.7\linewidth]{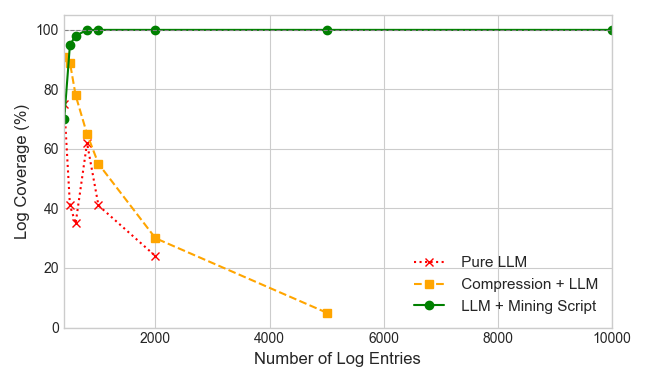} 
  \caption{Log Coverage (\%) vs. Log Size for the three evaluated methodologies.} 
  \label{fig:ablation_coverage}
\end{figure}

The results, presented in Table~\ref{tab:ablation_comparison} (Number of rules generated and Execution time) and visualized in Figure~\ref{fig:ablation_coverage} (Log coverage \%), are unambiguous. It is observed that our final hybrid approach is both significantly faster and, more importantly, is the only method that can reliably scale. While the LLM-only approaches fail to maintain high coverage on larger datasets, the performance of LANTERN improves with more data, consistently reaching 100\% coverage. This confirms that for a complex synthesis task like policy mining, the optimal architecture is an (LLM + Mining Script). Using deterministic algorithms for scalable, logical reasoning, and LLMs for well-defined language and scripting tasks is the best way forward, which answers research question RQ4.

\section{Related Work}
\label{sec:related}

Research on ABAC policy mining has evolved since its inception about a decade back \cite{stolleraccesslogmining}\cite{tdscstoller}, with some of the work finding their roots in the seminal efforts made for role mining in the context of RBAC \cite{roleminingccs}\cite{barshacsur}. Similar to role mining, the policy mining problem, i.e., Finding a minimal set of ABAC rules that satisfy a given collection of authorizations, has also been shown to be NP-complete. 
Das et al. \cite{gini}, for instance, present an algorithm that covers permissions in the given ACM based on Gini Impurity metric. The approach achieves high accuracy but often generates far from optimal number of ABAC rules. In contrast, semantically correct rules are generated by the method proposed by Batra et al. \cite{gunjantoit}. A visual means of mining ABAC policies has been proposed in \cite{vismap} where the authors show the step by step grouping of authorizations in an ACM. This makes the algorithm attractive from an implementation perspective as the system security officers can visualize the effect of policy mining and hence, can easily identify if unusual rules are getting generated. Another fairly recent work focuses on using hierarchical clustering and relationship extraction for policy mining \cite{shang2024abac}.

The feasibility of ABAC policy mining from a theoretical standpoint has been studied in \cite{sandhufeasibiliypolicymining}. The work identifies under what conditions policies can indeed be derived meaningfully from the inputs. It has also been identified that, the process of policy mining and access enforcement need to go together in a real ABAC system. Towards this, Talegaon et al. \cite{contemporaneous} propose a method that mines policies and also enforces access control contemporaneously. In an interesting related work, Madkaikar et al. \cite{madkaikarcose} show how the performance of ABAC systems can be measured using a queuing theoretic framework where a server normally mediates access requests. However, occasionally, it goes on \textit{vacation} to rebuild the policy from the current policy and an auxiliary list of specific authorizations. An M/G/1 queue with vacation model is used for analysis.

Besides Authorization Lists to ABAC rule generation as discussed above, efforts have been made towards obtaining ABAC rules from a policy written in a natural language. The goal is to convert such natural language access control policies to machine enforceable security policies~\cite{slank2014relation}. One of the approaches only extracts the attributes from NLACPs \cite{manar2} but not the complete policy. In contrast, ABAC rules are actually extracted in \cite{manar1}. However, it works for relatively simpler policies. Masoud et al. \cite{masoud3} extract complete policies from their natural language descriptions. Abdelgawad et al. \cite{indrakshisacmat23} go one step further by not only synthesizing, but also performing analysis of ABAC models extracted from NLACPs. 

It may be noted that generation of machine-enforceable ABAC rules from NLACPs using NLP techniques requires a very large corpus for training the models. Availability of such data is indeed a challenge in the access control domain. This problem has been eased to a great extent with the advent of powerful and versatile LLMs. Access control researchers have also started making use of these LLMs for generating MESPs from NLACPs \cite{ramnlacp}. Direct prompts and Retrieval Augmented Generation have been attempted by their authors. In order to have focused retrieval for the domain of access control, knowledge distillation has been found to be additionally helpful in a recent work by Yang et al. \cite{miansacmat2025}. To make the process of NLACP to MESP generation more accessible, Sonune et al. \cite{sonune2025lmntoolgeneratingmachine} have built a web-based tool where any user can upload an NLACP text and the corresponding MESP is generated on-the-fly. LANTERN makes a contribution that both supplements and complements these efforts, extending the current state of the art.

\section{Conclusion and Future Work}
\label{sec:concl}
In this paper, we have presented LANTERN - a framework for making access control policies more accessible. By integrating two stages of an LLM-powered toolkit, we show that both the technical overhead of coding and opacity of formal outputs can be reduced. While our implementation uses Gemini and a specific policy mining algorithm, the framework is flexible enough for other LLMs and different policy mining algorithms to be plugged in seamlessly. 


Despite the promising results, there are also a few limitations. For instance, the quality and completeness of the discovered ABAC rules are dependent on that of the input logs. The experiments used clean, synthetically generated logs. Performance on real-world, noisy, and incomplete log data may vary and would require additional pre-processing steps. 
Although our evaluation showed the miner performing well on tens of thousands of log entries, its performance on datasets with millions of entries has not yet been tested. 
LANTERN relies on the output of a large language model, which can be non-deterministic and is not guaranteed to be correct all the time. 
Notwithstanding the above shortcomings, we feel LANTERN makes an impactful first step towards converting logs into human comprehensible access control policies. Addressing the identified challenges would be our future direction of work.

    \bibliographystyle{splncs04}
\bibliography{sample-base}

\appendix
\section{Appendix - Illustrative Example}
\label{sec:appendix}


In this Appendix, we present an example user interaction with LANTERN, which begins with three key input format descriptions as shown in Listing \ref{lst:attribandlogformat}.
The LLM-based code generation module of LANTERN dynamically creates specialized parsing functions tailored to the user’s data schema. A part of the script generated by the LLM is shown in Listing \ref{lst:miningcode}. Listing \ref{lst:mininginput} illustrates the core data flow in policy mining. 
A sample output of the policy mining script is shown in Listing \ref{lst:minedrules}. As seen in the listing, each rule is an object that represents a general permission, formulated to maximize the coverage of the observed \textit{Allow} permissions in the logs (Listing \ref{lst:mininginput}) while maintaining conciseness through lower WSC. 
The summarizer is guided by a prompt that instructs it to act as a cybersecurity analyst and translate the technical rules into a professional, high-level executive summary. The goal is to distill the core business logic from the formal syntax. For the example rules shown in Listing~\ref{lst:minedrules}, the LLM would produce a natural language access control policy as shown in Listing~\ref{lst:summary_output}.

\begin{lstlisting}[style=SQLStyle, caption={Attributes and Log format},label={lst:attribandlogformat}]
user: <alice department:Finance designation:Manager>
object: <report.finance type:Financial sensitivity:High>
log: <alice report.finance read 2024-10-15 Allow>
\end{lstlisting}


\begin{lstlisting}[language=Python,style = CStyle, caption={Mining code snippet (Stage 1 output)}, label={lst:miningcode}]
class ABACPolicyMiner:
    def load_data(self, users, resources, logs):
        # Store parsed user, resource, and log data 
        pass
    def mine_abac_policy(self):
        # Example rule mining logic:
        # Identify patterns 
        # Have read and write permissions on resources
        # Use (WSC) metric for rule quality
        # Return list of mined rules with
        # attribute conditions and permissions
        pass
def parse_data_file(file_content: str, file_type: str):
    """
    Parses user, resource, or log data from raw text 
    input inferred from provided format examples.
    Extracts attribute key-value pairs. 
    Returns structured data as dicts/lists 
    """
    # Implementation details inferred by LLM...
    pass
\end{lstlisting}
\label{fig:miningcode}
\clearpage

\begin{lstlisting}[style = SQLStyle, caption={Example input to the mining script}, label={lst:mininginput}]
Input:
    Users Attribute:
    <morgan_finance_1 department:Finance designation:Manager>
    <taylor_sales_0 department:Sales designation:Manager>
    <alex_it_1 department:IT designation:System_Admin>
    <jordan_hr_0 department:HR designation:Generalist>
    Object Attribute:
    <financial_low_0.xlsx type:Financial sensitivity:Low>
    <financial_medium_0.xlsx type:Financial sensitivity:Medium>
    <financial_high_0.xlsx type:Financial sensitivity:High>
    <operational_low_0.csv type:Operational sensitivity:Low>
    <hr_medium_1.pdf type:HR sensitivity:Medium>
    Logs:
    <alex_it_1 hr_medium_1.pdf read 05:21:48 Allow>
    <jordan_hr_0 operational_low_0.csv read 05:22:33 Allow>
    <taylor_sales_0 operational_low_0.csv read 05:21:11 Allow>
    <morgan_finance_1 financial_low_0.xlsx read 05:21:46 Allow>
    <taylor_sales_0 financial_medium_0.xlsx read 05:26:05 Allow>
    <alex_it_1 financial_high_0.xlsx read 05:24:43 Allow>
\end{lstlisting}
\label{fig:mininginput}


\begin{lstlisting}[style=SQLStyle, caption={Example output of the policy mining script}, label={lst:minedrules}]
Output:
Rule 1: <User_Expr: {'designation': {'Manager', 'System_Admin'}, 'department': {'IT', 'Sales'}}, Resource_Expr: {'sensitivity': {'High', 'Medium', 'Low'}, 'type':  {'HR', 'Financial'}}, Operations: {'read'}, Constraints: set()>
  WSC (Complexity): 10

Rule 2: <User_Expr: {'department': {'HR', 'Sales'}, 'designation': {'Manager', 'Generalist'}}, Resource_Expr: {'type': {'Operational', 'Financial'}, 'sensitivity': {'Low'}}, Operations: {'read'}, Constraints: set()>
  WSC (Complexity): 8
\end{lstlisting}
\begin{lstlisting}[style=SQLStyle, caption={Natural language policy from the LLM summarizer (Stage 2 output)}, label={lst:summary_output}]
This document provides a high-level overview of our corporate Data Access Control Policy.
#Access Principles
* Broad Read Access for Management: A key pattern indicates that users with managerial roles (such as 'Manager' or 'System_Admin') in the IT and Sales departments have broad 'read' access to a wide range of sensitive data, including 'HR' and 'Financial' documents of all sensitivity levels.

* Focused Access for HR and Sales Staff: Another significant rule grants employees in the HR and Sales departments, including roles like 'Manager' and 'Generalist', 'read' access to low-sensitivity 'Operational' and 'Financial' data. This appears to correspond to standard operational duties.

Conclusion:
-----------
The mined policy reflects a clear distinction between the broad oversight permissions granted to management and the more focused, operational permissions granted to other staff. No complex cross-attribute constraints were discovered in this rule set.
\end{lstlisting}

\end{document}